\documentclass[aps,prl,twocolumn,showpacs]{revtex4}
\usepackage{color}
\usepackage{graphicx}
\usepackage{cancel}
\usepackage{dsfont}
\usepackage{bm}
\usepackage{dcolumn}

\def \tr      {\rm{tr}}

\begin{document}

\title{Duality of the spin and density dynamics for two-dimensional electrons with a spin-orbit coupling}

\author{I. V. Tokatly$^{1,3}$ and E. Ya. Sherman$^{2,3}$}
\affiliation{$^1$ Nano-bio Spectroscopy group and ETSF Scientific Development Centre, Dpto. F\'isica de Materiales, Universidad del Pa\'is Vasco, Centro de F\'isica de Materiale CSIC-UPV/EHU-MPC, E-20018 San Sebastian, Spain \\
$^2$Department of Physical Chemistry, Universidad del Pa\'is Vasco UPV-EHU,
48080 Bilbao, Spain \\
$^3$ Basque Foundation for Science IKERBASQUE, 48011, Bilbao, Spain}

\begin{abstract}
We study spin dynamics in a two-dimensional electron gas with a pure gauge
non-Abelian spin-orbit field, for which systems with balanced Rashba and
Dresselhaus spin-orbit couplings, and the (110)-axis grown GaAs quantum wells
are typical examples. We demonstrate the duality of the spin evolution and 
the electron density dynamics in a system without spin-orbit
coupling, which considerably simplifies and deepens the analysis of spin-dependent processes.
This duality opens a venue for the understanding of this class of systems, highly
interesting for their applications in spintronics, through known properties
of the systems without spin-orbit coupling.
\end{abstract}

\pacs{72.25.-b}

\maketitle

The understanding of spin dynamics in a two-dimensional (2D) electron gas with spin-orbit (SO) interaction is
highly important both for theoretical and applied spintronics, \cite{Winkler03,Zutic04,Dyakonov08,Wu10} including the design of devices 
with controlled spin transport. In many physically interesting situations the SO coupling can 
be elegantly described as an effective non-Abelian vector potential. \cite{Mineev92,Frolich93,Aleiner01,Levitov03,Lyanda,Yang06,Liu07,Natano07,Yang08,Leurs08,Raimondi,Tokatly08,Tokatly10}
There exists a class of systems, where the 
SO coupling corresponds to a pure gauge non-Abelian field. Therefore it can be 
gauged away and the behavior of a physical system should map to that of a system without SO coupling. In practice, 
there are at least two systems of this sort widely investigated from the applied point of view. 
These are 2D electrons with balanced Rashba and Dresselhaus couplings, and the electron gas 
in the (110)-axis grown GaAs quantum wells.

The coupled spin-charge dynamics is commonly described using the diffusion approximation,
where the rate of the spin precession is much less than
the momentum scattering rate.\cite{Dyakonov73,Diffusion}  
Currently, high-mobility 2D structures, where
the time scale of momentum relaxation is longer than
the spin rotation time,\cite{Griesbeck09,Korn10} became
available. Since here spins can make several turns between
collisions with impurities, the conventional Dyakonov-Perel mechanism \cite{Dyakonov73} 
is not applicable,\cite{Culcer07} and another type of analysis is required. In the present paper 
we solve this problem for systems with a pure gauge SO coupling, including quantum effects
due to the weak localization. 
We show for this general 
class of systems the existence of a duality of observables allowing the spin dynamics to be fully mapped
to the density dynamics in systems without SO coupling. As a result, several
regimes, including magnetic field dependence of the spin dynamics,
can easily be explored using a single formula.

We represent the Hamiltonian of a 2D electron gas with SO 
coupling as follows \cite{Tokatly08,Tokatly10} (the system of units with $\hbar=1$ is employed)
\begin{equation}
H=\frac{1}{2m}\int d^{2}{\rho}\Psi^{+}\left( i\partial _{i}+\mathcal{A}%
_{i}\right) ^{2}\Psi+W\left[\Psi^{+},\Psi\right],   \label{H}
\end{equation}
where $\Psi (\bm{\rho})$ is a spinor field operator, and the functional
$W\left[\Psi^{+},\Psi\right]$ contains all spin-independent contributions,
including  the external potential, electron-electron interactions, and, possibly, the electron-phonon coupling.
Here $m$ is the electron effective mass, and $\mathcal{A}_{i}$ are $2\times 2$
matrix-valued components of a non-Abelian $SU(2)$ gauge field describing the
SO coupling. In the broad class of systems of interest, $\mathcal{A}_{i}$ is a pure gauge, that is it
can be removed by a local $SU(2)$ transformation. The general form of a pure gauge vector potential,
\begin{equation}
\mathcal{A}_{i}=m\alpha_{i}\left( \mathbf{h}\cdot{\bm\sigma }\right) ,
\label{Aj}
\end{equation}
corresponds to the following SO Hamiltonian 
\begin{equation}
H_{\mathrm{so}}=\alpha \left( \mathbf{h}\cdot{\bm\sigma }\right) \left( \mathbf{k}\cdot{\bm\nu }\right) , 
 \label{Hso}
\end{equation}
where $\bm\sigma$ is a vector of Pauli matrices, $\alpha$ is the SO coupling constant, $\mathbf{h}$ is a
three-component unit vector for the SO field direction, ${\bm\nu }$ is a 2D vector in the $(x,y)$ plane, and $\alpha_{i}=\alpha \nu_{i}.$ 
The two practically important systems described by the Hamiltonian of Eq.~(\ref{Hso}) are: (i) the balanced Rashba-Dresselhaus system \cite{Averkiev99,Schliemann03} with $\mathbf{h=}\left( \pm 1,\pm 1,0\right) /%
\sqrt{2},$ ${\bm \nu }=\left( \pm 1,\pm 1\right) /\sqrt{2},$  and (ii) the $(110)$-axis
grown GaAs quantum well,\cite{Dyakonov86,Muller08,Belkov08,Tarasenko09} where $\mathbf{h}=\left(
0,0,1\right),$ ${\bm\nu}=\left(1,0\right) $ with the well axes
chosen with respect to the crystal axes as $x\parallel [1\overline{1}0],$ $%
y\parallel [001],$ and $z\parallel [110]$. Both systems are
expected to demonstrate highly anisotropic spin relaxation times with the spin
component along the $\mathbf{h}$-axis having a very low relaxation rate,
arising only due to a spin-dependent disorder.\cite{Spindisorder} 
Spin currents in the thermodynamical equilibrium state,\cite{Rashba03}
being common for 2D electron systems with SO coupling, are absent \cite{Tokatly08} in the 
structures described by the Hamiltonian in Eq.~(\ref{Hso}).

A local $SU(2)$ transformation, which gauges away the above type of
SO coupling is $\tilde{\Psi}(\bm{\rho})=\mathbf{U}_{\mathcal{A}}\Psi(\bm{\rho})$, where 
\begin{equation}
\mathbf{U}_{\mathcal{A}}=\exp \left[ {\rm i}m\alpha \left( \mathbf{h}\cdot{\bm\sigma}%
\right) \left( {\bm\rho}\cdot{\bm\nu}\right) \right].  \label{UA}
\end{equation}
The transformation (\ref{UA}) renders invariant all spin-independent
quantities, such as the charge and current densities, while the spin density
transforms covariantly: 
\begin{equation}
\widetilde{\mathbf{S}}=\frac{1}{2}\tr\{\bm\sigma\mathbf{U}_{\mathcal{A}%
}^{-1}(\mathbf{{S}\cdot\bm\sigma)\mathbf{U}_{\mathcal{A}}\}.}  
\label{Stilde}
\end{equation}
When the SO coupling is gauged away the dynamics of the transformed
spin density $\widetilde{\mathbf{S}}({\bm\rho},t)$ reduces to the spin
dynamics in the electron gas without SO interaction, which significantly
simplifies the analysis. Then, the physical spin density $\mathbf{S}({\bm\rho},t)$ 
is restored by 
\begin{equation}
\mathbf{S}=\frac{1}{2}\tr\{\bm\sigma\mathbf{U}_{\mathcal{A}}(\widetilde{%
\mathbf{S}}\cdot\bm\sigma)\mathbf{U}_{\mathcal{A}}^{-1}\},  \label{Sphysical}
\end{equation}
to obtain measurable results. Here we follow this guideline and show that
this approach allows to describe all regimes of spin dynamics on the same
footing.

We consider a 2D electron gas with a SO coupling of Eq.~(\ref{H}) and,
initially, a uniform spin density $\mathbf{S}$ produced, for example, by a
static magnetic field $\mathbf{B}$. At $t=0$ the magnetic field is released
and the spin relaxes due to a disorder potential and other interactions. To describe
this process we first eliminate the SO by the gauge transformation of Eq.~(\ref{UA}). 
Using Eq.~(\ref{Stilde}) we find that the initial uniform
physical spin density is mapped to the spin texture 
$\widetilde{\mathbf{S}}\left( {\bm \rho },0\right)=
\widetilde{\mathbf{S}}_{\parallel}\left( {\bm\rho},0\right)+\widetilde{\mathbf{S}}_{\perp }\left( {\bm \rho },0\right),$
where the term 
\begin{equation}
\widetilde{\mathbf{S}}_{\parallel}\left( {\bm\rho},0\right)=\mathbf{h}\left(\mathbf{S}\cdot\mathbf{h}\right),
\label{Stilde1}
\end{equation}
being parallel to $\mathbf{h}$, is untouched by the transformation and remains uniform, while the orthogonal 
to $\mathbf{h}$ part transforms into the helix structure \cite{Bernevig06,Koralek09} 
\begin{equation}
\widetilde{\mathbf{S}}_{\perp }\left( {\bm \rho },0\right) =\left[ \mathbf{%
S-h}\left( \mathbf{S}\cdot\mathbf{h}\right) \right] \cos (\mathbf{Q}_{\rm hx}\cdot{\bm\rho })-
(\mathbf{S}\times \mathbf{h})\sin (\mathbf{Q}_{\rm hx}\cdot{\bm\rho}),  
\label{helix1}
\end{equation}
where $\mathbf{Q}_{\rm hx}=2m\alpha {\bm\nu }$ is the helix wave vector.  

Since in the transformed system there is no SO coupling, the uniform part
of the initial spin distribution, Eq.~(\ref{Stilde1}),
is constant in time. A nontrivial dynamics occurs in the orthogonal
channel due to a diffusional decay of the initial helix spin texture
described by Eq.~(\ref{helix1}): 
\begin{equation}
\widetilde{S}_{\perp}^{\beta_1}\left( {\bm\rho},t\right) = \int\mathcal{D}%
^{\beta_{1}\beta_{2}}({\bm\rho}-{\bm\rho}^{\prime},t) \widetilde{S}%
_{\perp}^{\beta_2}\left( {\bm\rho}^{\prime},0\right)d^{2}{\bm\rho}^{\prime },
\label{S:evolution}
\end{equation}
where $\mathcal{D}^{\beta_{1}\beta_{2}}({\bm\rho}-{\bm\rho}^{\prime},t)$ is the exact spin diffusion
Green's function of a 2D electron gas, which takes into account the 
disorder, electron-electron and electron-phonon interactions. To proceed
further we note that in a nonmagnetic system without SO coupling the
spin diffusion Green's function is diagonal in spin subspace $\mathcal{D}%
^{\beta_{1}\beta_{2}}({\bm\rho},t)=\delta_{\beta_{1}\beta_{2}}\mathcal{D}({\bm\rho},t)$.
Hence Eq.~(\ref{S:evolution}) simplifies as 
\begin{equation}  \label{S:evolution1}
\widetilde{\mathbf{S}}_{\perp}\left( {\bm\rho},t\right) = \widetilde{\mathbf{%
S}}_{\perp}\left( {\bm\rho},0\right)\mathcal{D}(Q_{\rm hx},t),
\end{equation}
where $\mathcal{D}(q,t)$ is a Fourier component of the spin diffusion
Green's function 
\begin{equation}
\mathcal{D}(q,t) = \int d^2{\bm\rho}e^{-{\rm i}(\mathbf{q}\cdot{\bm\rho})}\mathcal{D}({\bm%
\rho},t),  \label{D:q}
\end{equation}
and we have taken into account that only the Fourier components of $\mathcal{D}({\bm\rho},t)$ with the
modulus of the wave vector $q=Q_{\rm hx}$ contribute to the dynamics of the
helix in Eq.~(\ref{helix1}).  Since the time-dependent factor in Eq.~(\ref{S:evolution1}) is scalar, the
transformation back to the physical spin, Eq.~(\ref{Sphysical}), simply reduces to removing tildas and the coordinate dependence in Eq.~(\ref{S:evolution1}). Thus, we get the following \textit{exact} result for the observable spin evolution 
\begin{equation}
\mathbf{S}_{\perp}\left(t\right)=\mathbf{S}_{\perp }(0)\int\frac{d\omega }{%
2\pi} \mathcal{D}\left( Q_{\rm hx},\omega \right)e^{-{\rm i}\omega t}.  
\label{S:perp}
\end{equation}
In Eq.~(\ref{S:perp}) we represented $\mathcal{D}(q,t)$ via the Fourier
integral because in the $\omega$-domain there is a simple expression of the
spin diffusion Green's function $\mathcal{D}(q,\omega)$ in terms of the
spin-spin correlator (the spin response function) $\chi_{\beta\beta}^{[S]}(q,\omega)$,
where $\beta=(x,y,z)$ is the Cartesian index corresponding to the spin component
\begin{equation}  \label{spin_response}
\mathcal{D}(Q_{\rm hx},\omega) = \frac{1}{{\rm i}\omega}\left[\frac{\chi_{\beta\beta}^{[S]}(Q_{\rm hx},%
\omega)}{\chi_{\beta\beta}^{[S]}(Q_{\rm hx},0)}-1\right].
\end{equation}
This equation can be derived by considering a linear response on a
time-dependent magnetic field that is adiabatically switched on at $t=-\infty$, and then suddenly switched off at $t=0$, i.~e. $\mathbf{B}(t)=e^{\delta t}\theta(-t)\mathbf{B}$ with $\delta\rightarrow 0$ (see, e.~g., Ref.~[\onlinecite{Belitz}] for similar calculations). Usually the SO coupling is weak on the Fermi energy scale, which implies $Q_{\rm hx}\ll k_F$, where $k_F$ is the Fermi momentum. 
Therefore in most situations one can safely replace the static $\omega=0$ response function 
in Eq.~(\ref{spin_response}) by the macroscopic Pauli spin susceptibility, 
$\chi_{\beta\beta}^{[S]}(Q_{\rm hx},0)=\chi_{\mathrm{P}}$.

Equations (\ref{S:perp}) and (\ref{spin_response}) are the result of 
the spin-density dynamics duality and give the \textit{exact}
evolution of the uniform spin
density. The problem is solved by mapping the spin relaxation in the
physical system to the ``washing out'' an inhomogeneous spin
texture in a dual system without SO coupling. The real spin relaxes because of SO-induced precession and
randomness introduced by disorder, phonons, and
interelectron interactions.\cite{eeinter} For the transformed spin, it is the
evolution of the nonuniform spin density distributions, again in the
presence of the disorder and interactions between the carriers. An
interesting exact feature of the pure gauge SO coupling (in addition to the
well known anisotropy) is the absence of the spin precession  
-- the vector $\mathbf{S}_{\perp}(t)$ is always collinear
to its initial direction. In the transformed picture this is related to the
diagonal structure of the spin response in a nonmagnetic electron gas. For
the real system this translates to the fact that spins of electrons with
opposite momenta precess around the $\mathbf{h}$-axis in the opposite
directions with the same rate. 

At the level of the random phase approximation the spin response function $%
\chi_{\beta\beta}^{[S]}(q,\omega)$ is equal to the density response function $%
\chi(q,\omega)$ of a noninteracting, but possibly disordered and/or coupled to
phonons electron gas, while the Pauli susceptibility $\chi_{\mathrm{P}}$ is proportional to the 
compressibility $\partial n/\partial \mu$, with $n$ and $\mu$ being
the electron concentration and chemical potential, respectively. Hence the spin diffusion Green's
function entering Eq.~(\ref{S:perp}) reduces to 
\begin{equation}
\mathcal{D}(Q_{\rm hx},\omega) = \frac{1}{{\rm i}\omega}\left[\frac{\chi(Q_{\rm hx},\omega)}{%
\partial n/\partial \mu} - 1\right],  
\label{D:Fourier}
\end{equation}
which is exactly the density diffusion Green's function. Therefore in this
physically important case the spin relaxation is mapped to the ordinary
density diffusion.

Now we apply Eqs.~(\ref{D:Fourier}) and (\ref{S:perp}) to a noninteracting disordered 2D
electron gas with a momentum relaxation time $\tau $, and study possible
regimes of spin dynamics. The density-density correlator $\chi(Q_{\rm hx},\omega)$
can be obtained either diagrammatically or by solving the kinetic equation.
In the semiclassical regime, corresponding to the summation of ladder
diagrams, one obtains: 
\begin{eqnarray}
\mathcal{D}\left( Q_{\rm hx},\omega \right) &=&\frac{\mathcal{K}\left(
Q_{\rm hx},\omega \right) }{1-\mathcal{K}\left( Q_{\rm hx},\omega \right) },
\label{ladder1} \\
\mathcal{K}\left( Q_{\rm hx},\omega \right) &=&\frac{1}{2\pi}\int \frac{d\theta 
}{1-{\rm i}\omega \tau +{\rm i}\Omega _{\mathrm{so}}\tau \cos \theta },  
\label{ladder2}
\end{eqnarray}
where the only SO-dependent parameter in the problem
$\Omega_{\mathrm{so}}\equiv Q_{\rm hx}v_{F}$ ($v_F$ is the Fermi velocity) 
is the maximum spin precession rate, 
and $\Omega_{\mathrm{so}}\tau=\ell Q_{\rm hx}$ (electron mean free path $\ell=v_F\tau$) 
characterizes the relaxation regime.

We begin with $\Omega_{\mathrm{so}}\tau\ll 1$ regime, which coresponds to a pure diffusion, 
studied in the coordinate representation in Ref.~[\onlinecite{Tokatly10}]. 
Here the diffusion Green's function, Eq.~(\ref{ladder1}), reduces to 
\begin{equation}
\mathcal{D}\left( Q_{\rm hx},\omega \right) =\frac{1}{DQ_{\rm hx}^{2}-{\rm i}\omega },
\label{diffuson}
\end{equation}
where $D=v_{F}^{2}\tau /2$ is the diffusion coefficient. Inserting $\mathcal{D}\left(Q_{\rm hx},\omega\right)$ of Eq.~(\ref{diffuson}) into Eq.~(\ref{S:perp}) we obtain:
\begin{equation}
\mathbf{S}_{\perp }\left( t\right) =\mathbf{S}_{\perp }(0)\exp \left(
-DQ_{\rm hx}^{2}t\right),\label{DP}
\end{equation}
which exactly corresponds to the Dyakonov-Perel' mechanism with the spin
relaxation rate $\Gamma _{s}=DQ_{\rm hx}^{2}$. Moreover, the factor $1/2$ in the
definition of $D$ acquires an interesting physical meaning in terms of the spin
precession: it corresponds to the angular averaging of the precession rate $%
\left\langle\Omega_{\mathrm{so}}^{2}(\mathbf{k})\right\rangle=\Omega_{%
\mathrm{so}}^{2}/2.$

The opposite, clean limit $\Omega _{\mathrm{so}}\tau \gg 1$, in terms of 
the dual (transformed) system corresponds to a reversible, purely ballistic washing out
the helix texture. In this regime $\mathcal{D}\left( Q_{\rm hx},\omega
\right)\approx\mathcal{K}\left( Q_{\rm hx},\omega \right)$ (see Ref.[\onlinecite{clean}]) 
and the integration in Eqs.~(\ref{S:perp}) and (\ref{ladder2}) yields: 
\begin{equation}
\mathbf{S}_{\perp }\left( t\right) =
\mathbf{S}_{\perp }(0)J_{0}\left(\Omega_{\mathrm{so}}t\right)
=\mathbf{S}_{\perp }(0)J_{0}\left(Q_{\rm hx}v_{F}t\right),  \label{S:perp:Bessel:1}
\end{equation}
where $J_{0}\left( \Omega _{\mathrm{so}}t\right) $ is the Bessel function.
The same result can be derived directly from the microscopic spin precession with a ${\bf k}$-dependent rate 
$\Omega_{\mathrm{so}}(\mathbf{k})=\Omega _{\mathrm{so}}\cos \phi$, where $\phi$ is the angle
between $\mathbf{k}$ and ${\bm\nu}$. Indeed, the net result of the inhomogeneous precession reproduces Eq.~(\ref{S:perp:Bessel:1}),
\begin{equation}
\mathbf{S}_{\perp }\left( t\right) =\mathbf{S}_{\perp }(0)\int \cos \left(
\Omega _{\mathrm{so}}t\cos \phi \right) \frac{d\phi }{2\pi }=\mathbf{S}%
_{\perp }(0)J_{0}\left( \Omega _{\mathrm{so}}t\right).
\end{equation}
In contrast, in the systems with pure Rashba or Dresselhaus coupling, spin precession rate does not
depend on the direction of momentum. As a result, in the ballistic regime, the $z$-component of the total
spin demonstrates cosine-like rather than the Bessel function time dependence.

\begin{figure}[t]
\includegraphics[width=6cm]{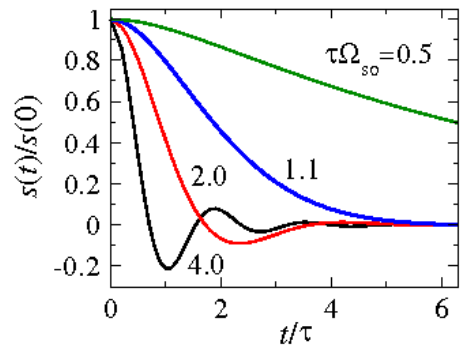}
\caption{(Color online.) Time dependence of the spin for different parameters of SO coupling, shown near the plots,
with $s(t)$ defined as $\mathbf{S}_{\perp}\left(t\right)\equiv s(t)\mathbf{S}_{\perp}\left(0\right)/{S}_{\perp}(0)$.}
\label{four_plots}
\end{figure}

An intermediate regime of $\Omega _{\mathrm{so}}\tau\sim1$, can be
investigated numerically. The results are presented in Fig.\ref{four_plots}
for different parameters. One can clearly see a crossover from the
oscillating Bessel function-like behavior to the exponential Dyakonov-Perel'
decay. At short times, the behavior of spin is universal: $\mathbf{S}_{\perp
}\left(t\right)=\mathbf{S}_{\perp }\left(0\right)\left(1-\Omega_{\mathrm{so}%
}^2t^2/2\right)$ due to the unperturbed precession of the spins. For the
density dynamics the universal short-time behavior is a direct consequence
of the $f$-sum rule,\cite{Vignale} as can be seen by expanding Eq.~(\ref{S:perp}) at $t\rightarrow 0$.

Analysis of Eqs.~(\ref{S:perp}),(\ref{ladder1}), and (\ref{ladder2}) shows
that $\Omega_{\mathrm{so}}\tau =1$ is a critical point. With the decrease in 
$\Omega _{\mathrm{so}}\tau$ in the range $\Omega _{\mathrm{so}}\tau >1,$
the first zero of $\mathbf{S}_{\perp}\left(t\right)$ rapidly shifts to
larger times, with negative value regions becoming very shallow. At $\Omega_{%
\mathrm{so}}\tau<1$ zeroes of $\mathbf{S}_{\perp }\left(t\right)$ disappear and the dynamics is a pure decay. 

The gauge transformation approach allows to analyze systems, where a direct treatment of the 
SO coupling would cause difficulties. The first effect we consider is the influence of the
orbital motion in a nonquantizing magnetic field $B$ along the $z$-axis on the spin dynamics.
We assume that due to a small $g-$factor of electron, the Zeeman coupling to the magnetic field 
does not cause a relevant spin precession. 
If $\Omega_{\rm so}\tau\ll 1$, the density evolution at $B=0$ is diffusive,  
and the electron mobility decreases with $B$ 
due to the Lorentz force as $\left(1+\omega _{c}^{2}\tau ^{2}\right) ^{-1}$,
where $\omega_{c}=|e|B/mc$ is the cyclotron frequency. 
By the Einstein relation, the diffusion coefficient
is renormalized by the same factor $D(B)=D(0)/\left(1+\omega_{c}^{2}\tau ^{2}\right)$. 
Hence the spin relaxation rate in Eq.~(\ref{DP}) decreases as 
$\Gamma_{s}(B)=\Gamma_{s}(0)/\left(1+\omega_{c}^{2}\tau^{2}\right)$, which reproduces the results of the direct quantum kinetic theory.\cite{Ivchenko73} 
For illustration, we consider the limit $\omega_{c}\tau\gg 1$ 
at short times $t\ll\tau$ in a more detail. 
Here the trajectories of electrons are very close to circles and the
spin-independent kernel in Eq.~(\ref{S:evolution}) can be represented as (cf. Ref.[\onlinecite{clean}])
\begin{equation}
\mathcal{D}({\bm\rho }^{\prime}-\mathbf{\bm\rho},t)=
\frac{1}{2\pi d(t)}\delta(\left|{\bm\rho}^{\prime}-{\bm\rho}\right|-d(t)),
\end{equation}
with the displacement $d(t)=2R_{c}\left|\sin\left(\omega_{c}t/2\right)\right|,$ 
where $R_{c}=v_{F}/\omega _{c}$ is the cyclotron
radius, and $R_{c}Q_{\rm hx}=\Omega_{\mathrm{so}}/\omega_{c}$. 
Straightforward integration in Eq.~(\ref{S:evolution}) yields (cf. Eq.(\ref{S:perp:Bessel:1}))
\begin{equation}
\mathbf{S}_{\perp}\left(t\right)=\mathbf{S}_{\perp}(0)J_{0}\left(Q_{\rm hx}d(t)\right),
\end{equation}
where $Q_{\rm hx}d(t)=2\Omega_{\mathrm{so}}\left|\sin({\omega_{c}t}/{2})\right|/{\omega_{c}}$.
For a very weak field (in a very clean system) $\omega_{c}\ll\Omega_{\mathrm{so}}$,  
the result reproduces Eq.~(\ref{S:perp:Bessel:1}), as
expected. In the opposite limit $\omega _{c}\gg \Omega _{\mathrm{so}}$, no
relaxation occurs. In terms of spin precession this can be understood\cite{Culcer07,Glazov07,Chang09}
as very fast changes at the frequency $\omega _{c}$ in the direction of the SO field, 
keeping the total spin out of relaxation. 
In terms of the nonuniform density dynamics,
this implies that the electrons, forced to circulate around small radius cyclotron orbits, can not spread out to destroy large scale $\sim 1/Q_{\rm{hx}}\gg R_{c}$ density variations. At long times, the diffusion behavior takes over, and the 
relaxation becomes exponential.  

As a second example we briefly discuss the effect of weak 
localization on the spin relaxation
\cite{Malshukov,Lyubinskiy04} by considering $\omega $-dependent renormalization 
of the momentum relaxation rate with the correction:\cite{Hirshfeld,Strinati,Dugaev86}
\begin{equation}
\delta\tau^{-1}_{\mathrm{wl}}(\omega)=\frac{2}{\tau}
\frac{1}{\partial n/\partial \mu}
\int\frac{d^{2}q}{\left(2\pi\right)^{2}}\frac{1}{Dq^{2}-{\rm i}\omega}.
\end{equation}
This correction arises due to the enhanced return probability, which slows down of the density 
dynamics and eventually leads to the algebraic tail in spin relaxation 
at long time as $\mathbf{S}_{\perp}\left(t\right) \sim 1/t.$ \cite{Lyubinskiy04}
In terms of spin precession, the enhanced backscattering slows down the spin relaxation because 
upon the return to the initial point the direction of the electron spin remains the same. 

To conclude, we have shown that in a wide class of systems
the non-Abelian gauge field description of SO coupling 
reveals the duality of experimental observables and ensures the
exact mapping of the spin dynamics to the
density evolution. The evolution is described in terms of the responce to an
external perturbation with the wavevector equal to the spin helix wavevector
$\mathbf{Q}_{\rm hx}$. We presented explicit results for a weak SO 
coupling with $Q_{\rm hx}\ll k_{F}$ valid for all systems of interest 
(this restriction is not required in general). 
This exact mapping opens a venue for understanding the whole class of practically important
systems through better studied, and more simple properties of the systems without SO coupling.

IVT acknowledges funding by the Spanish MEC (FIS2007-65702-C02-01), "Grupos
Consolidados UPV/EHU del Gobierno Vasco" (IT-319-07), and the European
Community through e-I3 ETSF project (Grant Agreement: 211956). This work of
EYS was supported by the University of Basque Country UPV/EHU grant GIU07/40,
MCI of Spain grant FIS2009-12773-C02-01, and "Grupos Consolidados UPV/EHU 
del Gobierno Vasco" grant IT-472-10.

\end{document}